\documentclass{article}
\usepackage{GEM_workshop_2025}
\usepackage{graphicx} 
\usepackage{appendix}
\usepackage{subcaption}
\usepackage{booktabs}
\usepackage{natbib}
\usepackage{amsmath}
\usepackage{amssymb}
\usepackage{comment}
\usepackage{siunitx}
\usepackage{textcomp}
\usepackage{hyperref}

\iclrfinalcopy

\title{Towards Interpretable Protein Structure Prediction with Sparse Autoencoders}

\author{Nithin Parsan \\
    Reticular \\
    \And
    David J. Yang\thanks{Work conducted during an internship at Reticular} \\
    University of Pennsylvania \\
    \And
    John J. Yang\thanks{Corresponding author. Contact \texttt{john@reticular.ai}} \\
    Reticular
}

\date{February 2025}

\begin{document}

\maketitle

\begin{abstract}

Protein language models have revolutionized structure prediction, but their nonlinear nature obscures how sequence representations inform structure prediction. While sparse autoencoders (SAEs) offer a path to interpretability here by learning linear representations in high-dimensional space, their application has been limited to smaller protein language models unable to perform structure prediction. In this work, we make two key advances: (1) we scale SAEs to ESM2-3B, the base model for ESMFold, enabling mechanistic interpretability of protein structure prediction for the first time, and (2) we adapt Matryoshka SAEs for protein language models, which learn hierarchically organized features by forcing nested groups of latents to reconstruct inputs independently. We demonstrate that our Matryoshka SAEs achieve comparable or better performance than standard architectures. Through comprehensive evaluations, we show that SAEs trained on ESM2-3B significantly outperform those trained on smaller models for both biological concept discovery and contact map prediction. Finally, we present an initial case study demonstrating how our approach enables targeted steering of ESMFold predictions, increasing structure solvent accessibility while fixing the input sequence. To facilitate further investigation by the broader community, we open-source our \href{https://github.com/johnyang101/reticular-sae}{code, dataset, pretrained models}, and \href{https://sae.reticular.ai}{visualizer}.

\end{abstract}

\section{Introduction}

Machine learning-based protein structure prediction models have achieved remarkable success by leveraging vast amounts of sequence data \citep{Jumper2021, Lin2022.07.20.500902}, building on the success of previous sequence homology-based methods \citep{marks2011couplings, gremlin2013}. However, their nonlinear nature obscures how sequence information informs structure prediction.

Previous interpretability studies on transformer-based protein language models (PLMs) have demonstrated PLMs learn structural information despite only training on sequences. \citet{rao2021transformer} and \citet{vig2021bertologymeetsbiologyinterpreting} showed that the attention map learns to predict residue-residue contacts. \citet{Ovc2024} show PLMs predict structure primarily by memorizing and retrieving patterns of coevolving residues. However, these works have yet to establish causal relationships between internal mechanisms and predictions.

Sparse autoencoders (SAEs) offer a promising direction for understanding language models by learning interpretable linear representations \citep{templeton2024scaling, gao2024scalingevaluatingsparseautoencoders}. Prior works \citep{Simon2024.11.14.623630, Adams2025.02.06.636901} train SAEs to uncover thousands of biologically interpretable features in ESM2 models, but these studies focused on smaller variants rather than ESM2-3B, the language model underlying ESMFold's structure prediction.

Our work advances protein model interpretability by scaling SAEs to ESM2-3B (the base model for ESMFold) and also adapting Matryoshka SAEs \citep{nabeshima2024matryoshka, bussmann2024matryoshka} to learn hierarchically organized features that align with protein structure's multi-scale nature.

We structure the paper as follows: In Section 2, we present scaling SAEs to ESM2-3B and Matryoshka SAEs for hierarchical features. Section 3 shows Matryoshka SAEs match or exceed L1 and TopK in language modeling and structure prediction. Section 4 evaluates ESM2-3B's performance on biological concept discovery and contact prediction. Section 5 presents feature steering to control protein structure properties. Code, models, and a visualizer will be released upon publication.


\section{Methods}
\begin{figure}
    \centering
    \includegraphics[width=0.9\linewidth]{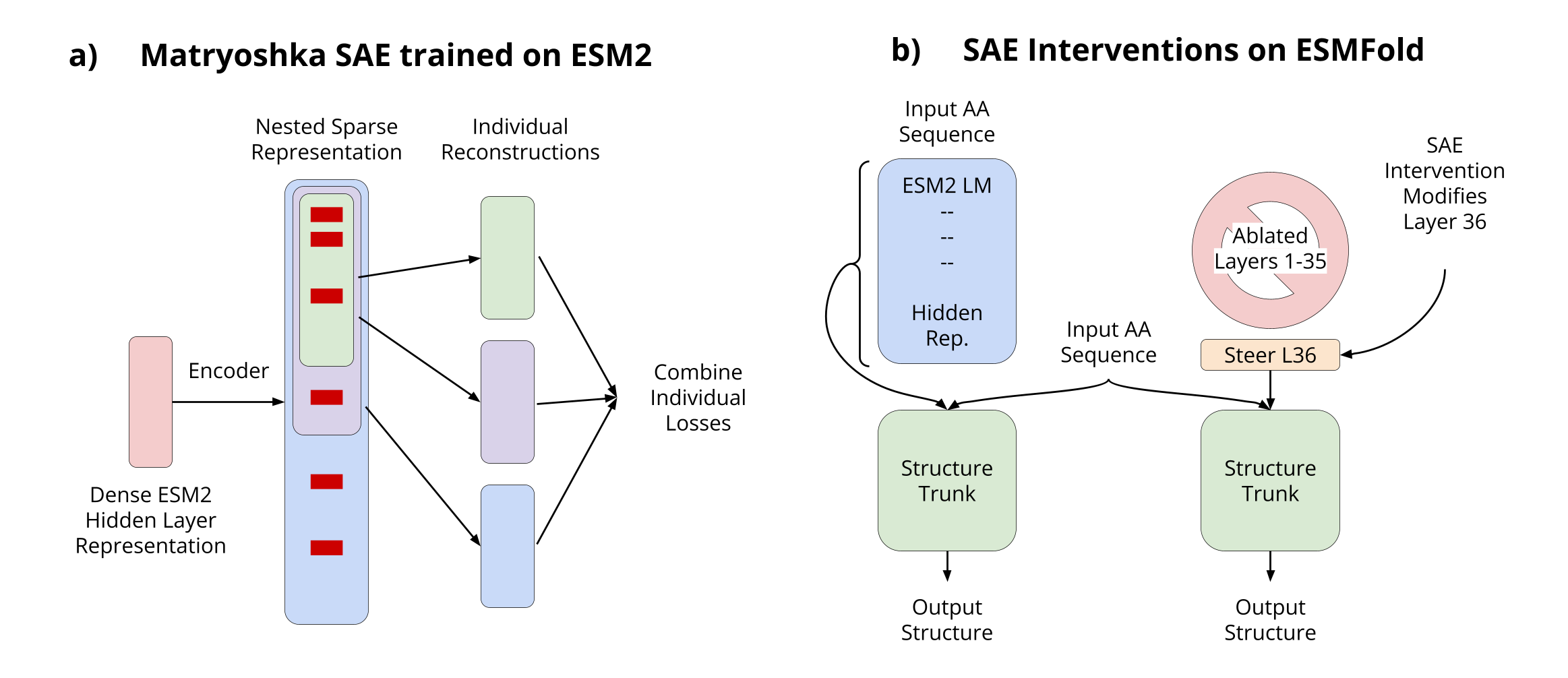}
    \caption{a) Matryoshka Sparse Autoencoder (SAE) architecture for training on ESM2 hidden layer representations, showing nested sparse feature organization. b) SAE intervention framework for ESMFold, comparing normal operation (left) where all ESM2 hidden representations flow to the structure trunk, versus intervention (right) where only a modified layer 36 representation is used while ablating all other layers.}
    \label{fig:diagram}
\end{figure}
\subsection{Setup} 
\textbf{Problem Statement.} Given token embeddings $x \in \mathbb{R}^d$ from a transformer-based PLM, our SAE encodes $x$ into a sparse, higher-dimensional latent representation $z \in \mathbb{R}^n$ where $d << n$, and decodes it to reconstruct $x$ by minimizing the L2 loss $\mathcal{L} = \left| x - \hat{x} \right|^2_2$. We enforce sparsity on $z$ through methods detailed in Appendix \ref{app:sae_details}.

\textbf{Models and Training.} We train SAEs on layers 18 and 36 of ESM2 \citep{Lin2022.07.20.500902}, focusing on the 3 billion parameter ESM2-3B model which provides representations for ESMFold. During training, embeddings are normalized to enable hyperparameter transfer between layers, with biases scaled during inference following \citet{marks2024dictionary_learning}. Our training data consists of 10M sequences randomly selected from UniRef50, constituting 2.5 billion tokens. See Appendix \ref{app:data_curation} for details.
\subsection{Matryoshka SAEs}
Proteins exhibit inherent hierarchical organization across scales, from local amino acid patterns to molecular assemblies. We employ Matryoshka Sparse Autoencoders (SAEs), which learn nested hierarchical representations through embedded features of increasing dimensionality (Fig. \ref{fig:diagram}a) \citep{nabeshima2024matryoshka, bussmann2024matryoshka}. 

Our implementation follows \citet{bussmann2024matryoshka}, \citet{marks2024dictionary_learning}, and \citet{gao2024scalingevaluatingsparseautoencoders} in dividing the latent dictionary into nested groups where earlier groups capture abstract features while later groups encode granular details. For mathematical details of the encoding, decoding, and loss computation, see Appendix \ref{app:matk_math}.

\begin{figure}
    \centering
    \begin{subfigure}{0.45\linewidth}
        \centering
        \includegraphics[width=\linewidth]{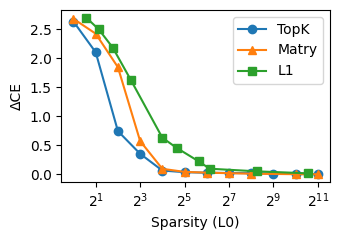}
        \caption{CE loss reconstruction across sparsity levels for TopK, Matryoshka (Matry), and L1 regularized autoencoders. Matryoshka requires similar or less active latents to achieve good reconstruction.}
        \label{fig:delta_ce}
    \end{subfigure}
    \hfill
    \begin{subfigure}{0.45\linewidth}
        \centering
        \vspace{0.5cm} 
        \resizebox{\linewidth}{!}{%
        \begin{tabular}{lc}
            \toprule
            Comparison & RMSD (Å) \\
            \midrule
            Exp vs. ESMFold (baseline) & 3.1 ± 2.5 \\
            Exp vs. ESMFold (full ablation) & 15.1 ± 5.9 \\
            Exp vs. ESMFold (only layer 36) & 2.9 ± 2.1 \\
            Exp vs. SAE (layer 36) & 3.2 ± 2.6 \\
            ESMFold vs. SAE (both L36) & 3.1 ± 4.4 \\
            \bottomrule
        \end{tabular}
        }
        \vspace{1.5cm} 
        \caption{Backbone RMSD (Å) comparing experimental structures (Exp), ESMFold predictions, and SAE reconstructions. Keeping layer 36 or using SAE preserves accuracy, while full ablation degrades performance.}
        \label{fig:casp14_table}
    \end{subfigure}
\end{figure}

\section{Evaluations on Downstream Loss}

We evaluate our SAEs on language modeling and structure prediction for ESM2-3B and ESMFold respectively to assess how well they preserve performance on downstream tasks.

\subsection{Language Modeling}
\textbf{Setup.} We report the average difference in cross-entropy loss $\Delta CE$ between the logits from the original and SAE reconstructed PLM. We evaluate on a heldout test set of 10k sequences randomly sampled from Uniref50 on layer 18 of ESM2. 

\textbf{Results.} Figure \ref{fig:delta_ce} shows the impact of different autoencoder architectures on downstream language modeling performance across sparsity levels. For architecture details, see Appendix \ref{app:impl}. At low sparsity ($L_0 < 10$), all approaches - TopK, Matryoshka, and L1 regularization - show similar degradation ($\Delta \text{CE} \approx 2.5\text{--}3.0$). As sparsity increases, TopK and Matryoshka SAEs maintain better performance compared to L1 regularization, with $\Delta \text{CE}$ approaching 0 for sparsity levels above 100.

\subsection{Structure Prediction} 

\textbf{Setup.} By scaling SAEs to ESM2-3B, we can now evaluate how well our SAE representations preserve ESMFold's structure prediction capabilities. We focus on reconstructing ESM2's hidden representations that feed into ESMFold. Since ESMFold uses representations from all layers but our SAE reconstructs only one layer, we ablate all other layers to isolate reconstruction effects. This ablation maintains performance on the CASP14 test set (see Fig. \ref{fig:casp14_table}).

\textbf{Data.} We evaluate structure prediction on the CASP14 dataset, a diverse challenging benchmark held out from ESMFold during training. After filtering for computational constraints and prediction quality (see Appendix \ref{app:casp14_dataset}), we analyze 17 protein targets. 


\textbf{Results.} Our experiments demonstrate effective preservation of structural information. In Fig. \ref{fig:casp14_table}, we see that comparing experimental structures to ESMFold predictions shows an RMSD of 3.1 ± 2.5 Å without ablation. While full ablation of ESM2 embeddings significantly degrades performance (RMSD 15.1 ± 5.9 Å), both keeping only layer 36 (RMSD 2.9 ± 2.1 Å) and using our SAE reconstruction (RMSD 3.2 ± 2.6 Å) maintain comparable performance. The similarity between ESMFold and SAE predictions (RMSD 3.1 ± 4.4 Å) confirms preservation of structural information. We also see that in Fig. \ref{fig:rmsd_k_sweep}, with only 8 to 32 active latents, SAEs can reasonably recover structure prediction performance.

These results show our SAE effectively compresses model representations while maintaining both sequence-level and structural prediction capabilities across sparsity levels.

\section{Further Evaluations}

\subsection{Swiss-Prot Concept Discovery}
We evaluate feature interpretability through alignment with Swiss-Prot annotations following \citet{Simon2024.11.14.623630}'s methodology (details in Appendix \ref{app:swiss_prot_background}), analyzing 30,871,402 amino acid tokens across 476 biological concepts. Features capturing concepts (F1 > 0.5) are identified using domain-level recall with post-hoc [0,1] activation normalization across architectural variants.

Comparing ESM2-3B versus ESM2-8M models (dictionary size 20,480, sparsity k=100), 3B variants substantially outperform their 8M counterparts. The 3B Matryoshka and TopK models identify 233 concepts (48.9\%) with F1 $>$ 0.5, generating 2,677 and 2,461 high-quality feature-concept pairs respectively. In contrast, 8M Matryoshka and TopK variants capture only 72 (15.1\%) and 95 (20.0\%) concepts, with 287 and 844 pairs respectively. Performance gains are particularly pronounced in protein domains (76\% versus 19.5-28.1\% coverage). Head-to-head comparisons show 3B models achieve higher F1 scores on over 400 concepts (mean improvement 0.25), while architectural differences within each scale remain minimal (Figure \ref{fig:f1_heatmap}, details in Appendix \ref{app:swiss_prot_performance}).

\subsection{Contact Map Prediction}
\textbf{Background.} Following \citet{Ovc2024}, we evaluate our SAEs' ability to capture coevolutionary statistics through contact map prediction using the Categorical Jacobian. Details in Appendix \ref{app:cm}. This provides an unsupervised test of whether our compressed representations preserve the structural information encoded in the original model.

\begin{figure}
    \centering
    \begin{subfigure}{0.45\linewidth}
        \centering
        \includegraphics[width=\linewidth]{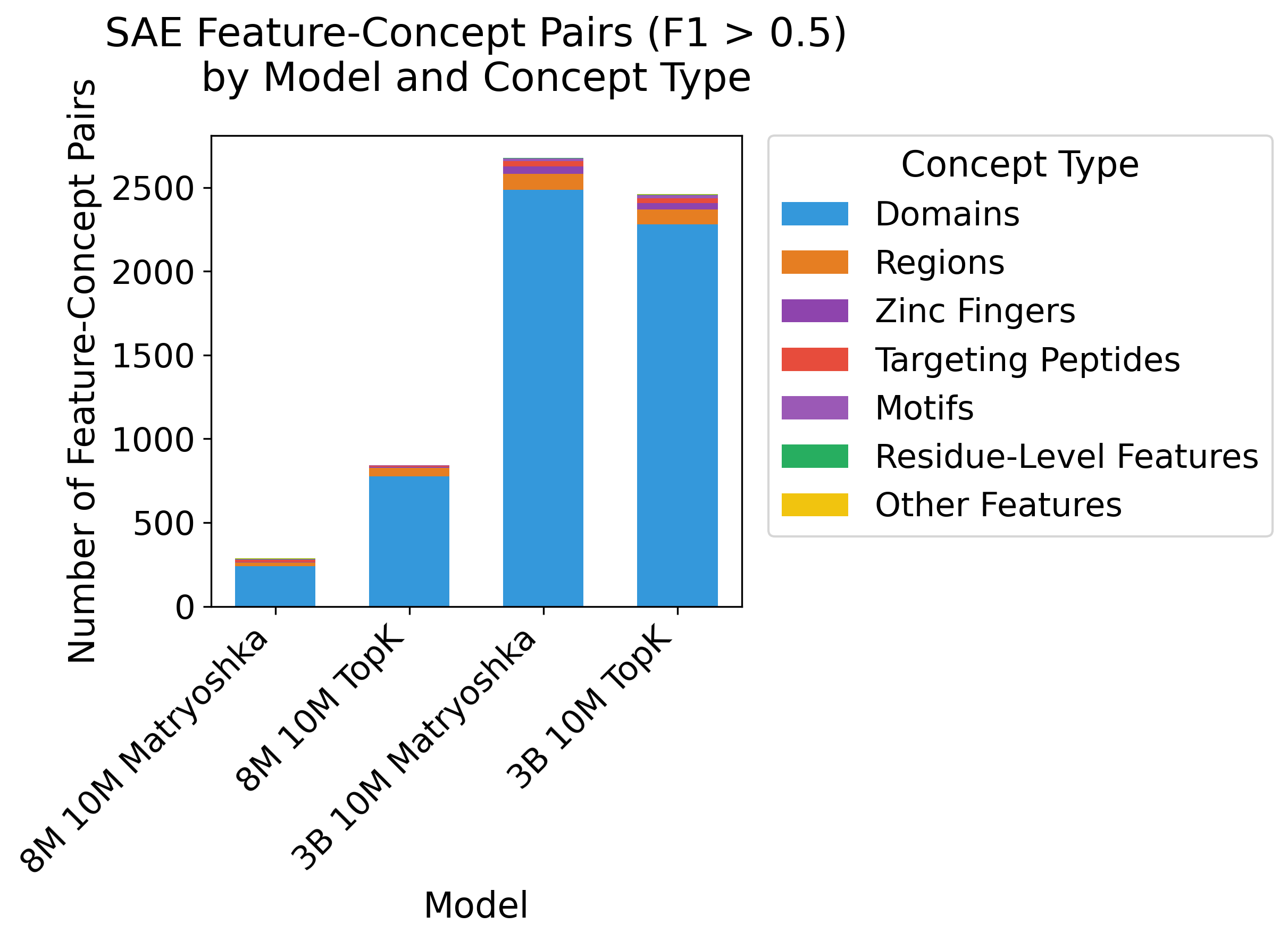}
        \caption{Number of high-performing feature-concept pairs (F1 $>$ 0.5) across model scales and architectures, broken down by concept type.}
        \label{fig:f1_concept_pairs}
    \end{subfigure}
    \hfill
    \begin{subfigure}{0.45\linewidth}
        \centering
        \includegraphics[width=\linewidth]{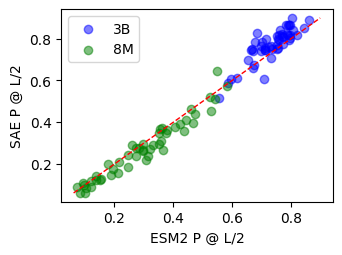}
        \caption{Correlation plot on long-range contact accuracy measured by Precision at L/2 (P @ L/2) between ESM2 and SAE reconstructions on 8M and 3B subject model sizes.}
        \label{fig:cm_plot}
    \end{subfigure}
    \caption{Analysis of feature-concept relationships and long-range contact accuracy.}
    \label{fig:concept_analysis}
\end{figure}

\textbf{Results.} We assess contact prediction accuracy using the precision at L/2 metric (P @ L/2), which measures the fraction of correctly predicted contacts among the top L/2 predicted long-range contacts, where L is the protein sequence length. Figure \ref{fig:cm_plot} shows the correlation between contact prediction accuracy of ESM2 and our SAE reconstructions across different model scales. The 3B model demonstrates consistently higher precision compared to the 8M model, with reconstructions closely tracking the original ESM2 predictions. This suggests that larger language models capture more robust coevolutionary signals that can be effectively compressed by our approach.

\section{Case Study: SAE Feature Steering on ESMFold}

\subsection{SAEs on ESMFold Help Identify Structural Features}
\begin{figure}[h]
    \centering
    \begin{subfigure}{0.60\linewidth}
        \centering
        \includegraphics[width=\linewidth]{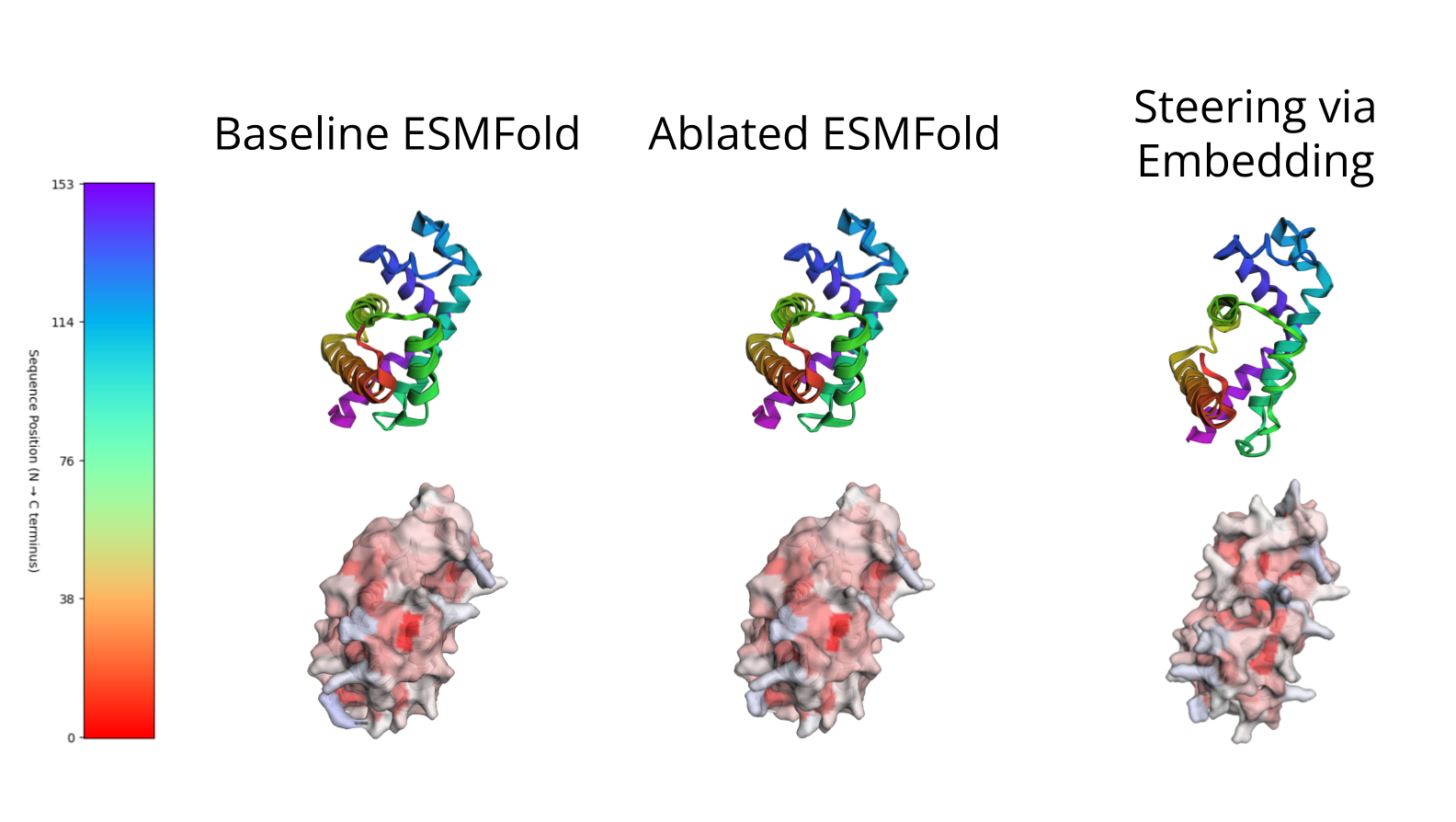}
        \vspace{0.5cm}
        \caption{Structural visualization of feature steering effect on myoglobin with $\alpha$= -0.275 on selected feature. Bottom row surface representation is colored by computed SASA, with blue as low, white as medium, and red as higher SASA.}
        \label{fig:case_study}
    \end{subfigure}
    \hfill
    \begin{subfigure}{0.35\linewidth}
        \centering
        \includegraphics[width=\linewidth]{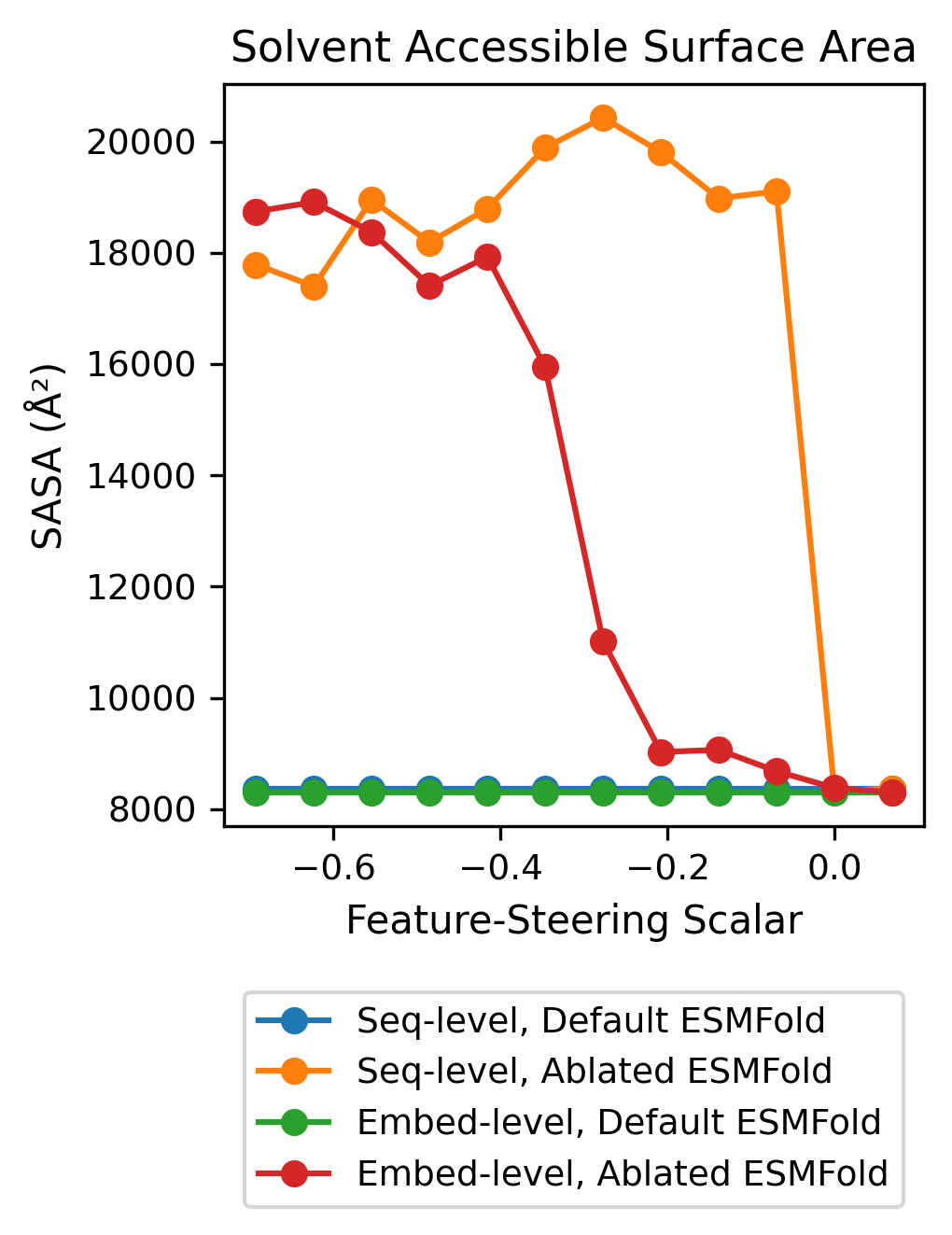}
        \caption{Solvent accessible surface area (SASA) changes under different steering conditions.}
        \label{fig:sasa_plot}
    \end{subfigure}
    \caption{Feature steering and SASA analysis.}
    \label{fig:steering_analysis}
\end{figure}

Building on previous work by \cite{Simon2024.11.14.623630}, we investigated whether targeted feature manipulation of ESM2-3B could induce coordinated changes in both sequence composition and predicted structure through ESMFold's pipeline. Using our Matryoshka SAE trained on ESM2-3B, we identified a feature strongly correlated with residue hydrophobicity. Following the feature steering methodology of \cite{templeton2024scaling} and detailed in Appendix Section \ref{app:case_study}, we steer this feature ($\alpha$ = -0.275) and observe significant changes in predicted protein surface accessibility of myoglobin (PDB ID: 1MBN) while maintaining structural integrity (Figure \ref{fig:case_study}).

Steering increased total Solvent Accessible Surface Area (SASA) by 31.5\% (from 8,369.5 to 11,009.3 \si{\angstrom\squared}) with minimal structural disruption (RMSD 2.76 \si{\angstrom}) in the expected range of Figure \ref{fig:casp14_table}. Feature intervention affected both sequence-level predictions via steering toward more hydrophilic residues and direct structural predictions of ESMFold (Figure \ref{fig:gravy_seq_similarity}). Critically, steering the hidden representation at layer 36 alone was sufficient to induce significant structural changes consistent with increased hydrophilicity, even when providing ESMFold with the correct input sequence (Figure \ref{fig:sasa_plot}). Further details found in Appendix Section \ref{app:case_study_details} and feature validation in Appendix Section \ref{app:case_study_controls}. 

\section{Discussion}

This work advances PLM interpretability by scaling SAEs to ESM2-3B, extending recent work using SAEs to interpret PLMs to the structure prediction task (\cite{Simon2024.11.14.623630}; \cite{Adams2025.02.06.636901}). Through a combination of increasing subject model size, leveraging the Matryoshka architecture, and targeted interventions on ESMFold's structure predictions, we present the first work applying mechanistic interpretability to protein structure prediction. A key limitation is that our focus is exclusively on how PLM-derived sequence representations inform structure prediction. Looking ahead, we aim to establish formal connections between SAE sequence representations and equivariant representations of geometric data. Additional limitations and future directions can be found in Appendix \ref{app:limitations} and \ref{app:future_work}, respectively.

\newpage
\section{Appendix}
\appendix

\section{Related Work} \label{app:related_work}
Recent work has focused on mechanistically interpreting protein language models (PLMs) to understand how they achieve remarkable success in protein modeling and design. Early interpretability studies examined PLM internals, with \cite{vig2021bertologymeetsbiologyinterpreting} and \cite{rao2021transformer} discovering that attention patterns encode structural relationships between amino acids. Later work extended these findings, showing attention could identify functionally important regions like allosteric sites (\cite{Kannan2024.10.03.616547}; \cite{Dong2024.09.28.615583}). 

Building on this attention-based analysis, \cite{Ovc2024} provided key mechanistic insights by showing that PLMs primarily learn by storing and looking up coevolutionary patterns preserved through evolution, rather than learning fundamental protein physics. By analyzing what sequence features influence contact predictions through a "categorical Jacobian" method, they demonstrated that PLMs memorize statistics of co-evolving residues analogous to classic evolutionary models.

Sparse autoencoders (SAEs) have emerged as a powerful new tool for interpreting these models. \cite{Simon2024.11.14.623630} developed InterPLM, extracting thousands of interpretable features from ESM-2 that correspond to known biological concepts like binding sites, structural motifs, and functional domains. Their work revealed that most biological concepts exist in superposition within the model's neurons, as SAEs vastly outperformed individual neurons at capturing these concepts. Concurrently, \cite{Adams2025.02.06.636901} explored different SAE training approaches and evaluated their biological relevance through novel downstream tasks. In this work, we scale SAE training and evaluation to ESM2-3B and introduce Matryoshka SAEs (\cite{bussmann2024matryoshka}; \cite{nabeshima2024matryoshka}) with hierarchical encoding to steer ESMFold's structure prediction capabilities for the first time.

\section{Additional Limitations} \label{app:limitations}
Below we outline additional limitations of our work. First, as mentioned earlier, although we take an important step toward mechanistic interpretability of protein structure prediction, our focus is exclusively on how PLM-derived sequence representations inform structure prediction. The interpretability of ESMFold's structure prediction head is left for future investigation. Second, our interventions on ESMFold were restricted to single-feature manipulations on embeddings from individual layers rather than exploring combinations of features or cross-layer interactions. Third, our analysis is limited to the 8M and 3B models of ESM2; evaluating a broader range of model sizes could reveal whether SAE performance eventually plateaus. Fourth, ESMFold is no longer state-of-the-art compared to newer diffusion-based protein structure prediction methods  \cite{Abramson2024, WohlwendBoltz124, Watson2023}. 

Fifth, recent work suggests that SAEs can learn different features even when trained on the same data, implying that our results may be sensitive to both initialization and the specific UniRef50 sample used (\cite{paulo2025sparseautoencoderstraineddata}). Sixth, our structural steering experiments focused primarily on surface accessibility, leaving other potential structural properties unexplored, and the observed hydrophobicity effects might be achievable through simpler approaches such as steering smaller PLMs or using supervised linear probes. Seventh, for Matryoshka SAEs, we do not report how the number of groups and group size choices impact results. Finally, although our ablation studies demonstrate preservation of structural information, we have not fully characterized how interactions among SAE features might affect ESMFold’s folding mechanism. We anticipate that future work leveraging our open-source models developed with substantial computational resources will address these limitations and yield further insights for the community.

\section{Future Work} \label{app:future_work}
\textbf{Connections to Geometric Representations.} This work, while taking a significant step towards interpreting protein structure prediction, focuses primarily on learning linear representations of sequence information. Future work could establish formal connections between SAE sequence representations and equivariant representations of geometric data \citep{ThomasTFN18, geiger2022e3nneuclideanneuralnetworks, LeeEquifold22}, particularly investigating how the linear overcomplete basis learned by SAEs map to irreducible representations of geometric transformations in the context of protein structure prediction.

\textbf{Theoretical Analysis of Matryoshka SAEs.} Additionally, theoretical analysis drawing from ordered autoencoding \citep{rippel2014learningorderedrepresentationsnested, xu21anytime} and information bottleneck methods \citep{tishby2000informationbottleneckmethod} applied to protein structure prediction could improve our understanding of multi-scale feature learning in biological sequences. Such analysis may provide deeper insights into how hierarchical feature representations emerge and interact within the context of protein language models.

\section{Matryoshka SAE Implementation Details} \label{app:matk_math}
The Matryoshka SAE architecture builds upon previous ones through nested groups of increasing size. This progressive constraint naturally encourages the emergence of a feature hierarchy - earlier groups must capture high-level, abstract features to achieve reasonable reconstruction with limited capacity, while later groups can encode more granular details. The key innovation lies in their group-wise decoding process, where each group must reconstruct the input using only its allocated subset of latents.

When processing a protein token embedding $x \in \mathbb{R}^d$ from a PLM, the encoding process follows as:
\begin{align}
z &= \text{BatchTopK}(W_{\text{enc}}x + b_{\text{enc}})
\end{align}
where $W_{\text{enc}} \in \mathbb{R}^{n \times d}$ is the encoder weight matrix, $b_{\text{enc}} \in \mathbb{R}^n$ is the encoder bias, and BatchTopK enforces sparsity by keeping only the $B * K$ highest activations in each batch. This enforces average sparsity while allowing the number of active latents per token to vary.

The decoding process is:
\begin{align}
\hat{x}^{(m)} &= W_{\text{dec}}^{(m)} z_{[1:m]} + b_{\text{dec}}
\end{align}
Here, $z_{[1:m]}$ represents the first $m$ components of the latent vector, $W_{\text{dec}}^{(m)} \in \mathbb{R}^{d \times m}$ is the decoder weight matrix restricted to its first $m$ columns, and $b_{\text{dec}} \in \mathbb{R}^d$ is the decoder bias. Each group $m$ must reconstruct the input using only its allocated subset of latents. The total loss combines reconstruction errors across all group sizes $M$:
\begin{align}
\mathcal{L}(x) &= \sum_{m \in M} \Vert x - \hat{x}^{(m)} \Vert_2^2 + \alpha \mathcal{L}_{\textrm{aux}}
\end{align}
where $\alpha$ weights any auxiliary regularization terms. We use the auxiliary loss term from \citet{marks2024dictionary_learning} and \citet{gao2024scalingevaluatingsparseautoencoders} to reduce dead latents.

\section{Swiss-Prot Evaluation} \label{app:swiss_prot_eval}

\subsection{Background of Swiss-Prot Evaluation} \label{app:swiss_prot_background}
Swiss-Prot is the gold standard manually curated section of the UniProt Knowledge Base, containing expert-annotated protein sequences with detailed information about their structure, function, modifications, and key biological regions (\cite{swissprotcitation}; \cite{uniprotkbcitation}). Through extensive experimental validation and literature review, Swiss-Prot provides high-quality annotations that span from individual catalytic sites to complete functional domains. Following \citet{Simon2024.11.14.623630}, we evaluate whether learned features align with these known biological concepts using a modified F1 score that handles the mismatch between precise feature activations and broader protein annotations present in Swiss-Prot. This approach calculates precision at the amino acid level but recall at the domain level, enabling systematic comparison between learned features and known biological concepts. 

\subsection{Performance of SAE Models on F1 Concept Evaluation} \label{app:swiss_prot_performance}

\textbf{Model Scale vs Architecture:} At 3B scale, architectural differences have minimal impact, with both Matryoshka and TopK identifying 233 concepts and showing only small variations in high-quality feature-concept pairs (2,677 vs 2,461). At 8M scale, architectural differences are more pronounced though still modest, with TopK outperforming Matryoshka (95 vs 72 concepts). However, the scale effect dominates these architectural differences - both 3B variants substantially outperform both 8M variants across all concept types. 


\begin{table}[htbp]
  \centering
  \begin{tabular}{lccccc}
    \toprule
    Category & Total Concepts & \multicolumn{4}{c}{Concepts with F1 $>$ 0.5 (\%)} \\
    \cmidrule(lr){3-6}
    & & 3B MatK & 3B TopK & 8M MatK & 8M TopK \\
    \midrule
    Domains & 256 & 193 (75.4\%) & 195 (76.2\%) & 50 (19.5\%) & 72 (28.1\%) \\
    Regions & 55 & 19 (34.5\%) & 20 (36.4\%) & 10 (18.2\%) & 13 (23.6\%) \\
    Motifs & 37 & 6 (16.2\%) & 5 (13.5\%) & 2 (5.4\%) & 3 (8.1\%) \\
    Zinc Fingers & 15 & 5 (33.3\%) & 5 (33.3\%) & 2 (13.3\%) & 2 (13.3\%) \\
    Other Features & 10 & 5 (50.0\%) & 3 (30.0\%) & 4 (40.0\%) & 1 (10.0\%) \\
    Targeting Peptides & 5 & 5 (100.0\%) & 5 (100.0\%) & 4 (80.0\%) & 4 (80.0\%) \\
    \midrule
    Total & 476 & 233 (48.9\%) & 233 (48.9\%) & 72 (15.1\%) & 95 (20.0\%) \\
    \bottomrule
  \end{tabular}
  \caption{Concept coverage comparison across model scales and architectures. We report the number (percentage) of concepts with F1 scores exceeding 0.5 for each model variant, broken down by concept category.}
  \label{tab:concept_coverage}
\end{table}

\begin{figure}[h]
    \centering
    \includegraphics[width=0.9\linewidth]{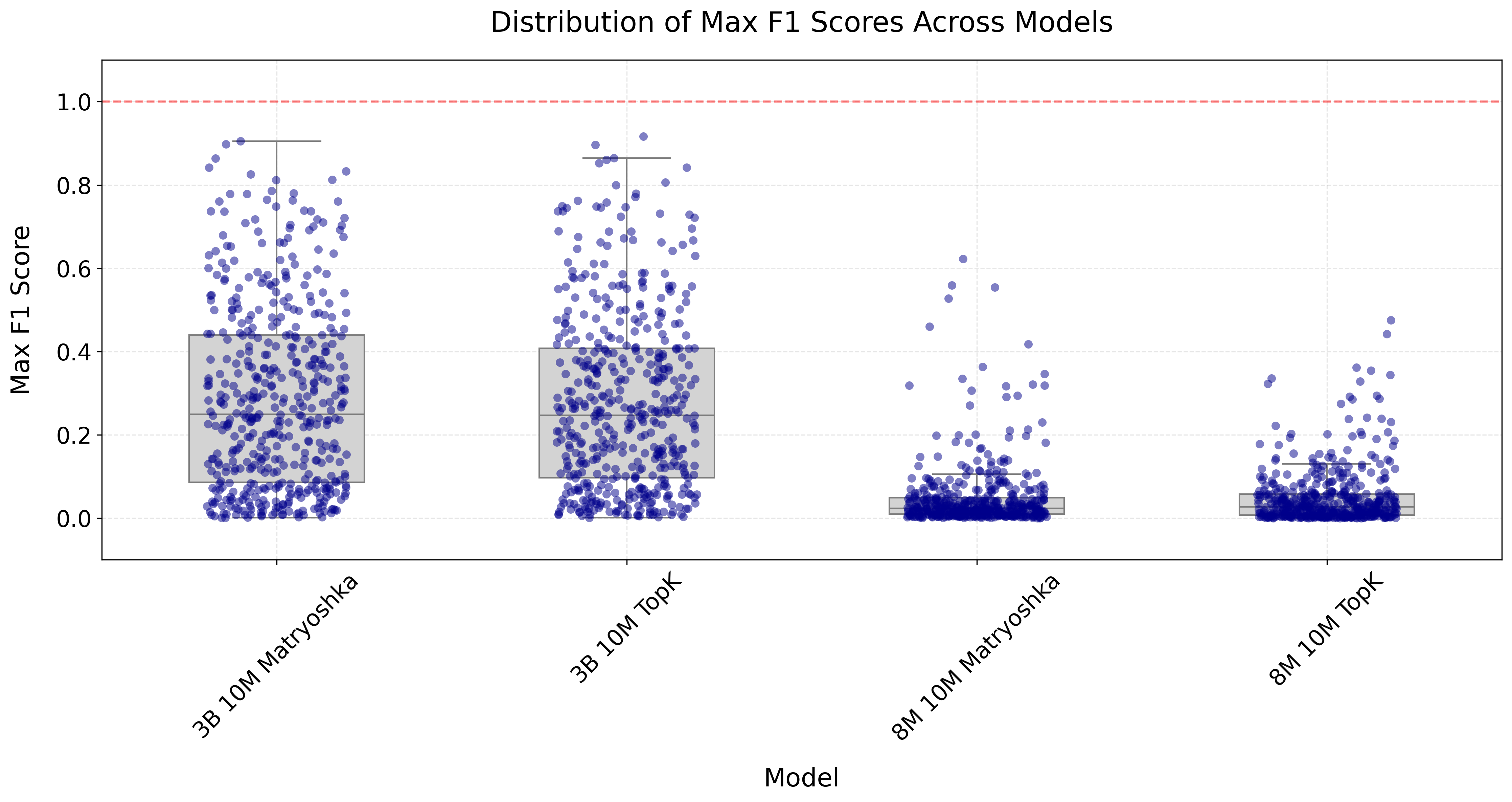}
    \caption{Distribution of highest F1 scores achieved for each concept across models.}
    \label{fig:f1_scores_dist}
\end{figure}

The boxplot distribution reveals a clear separation between model scales (\ref{fig:f1_scores_dist}. The 3B models show broader distributions with medians around 0.4 and numerous concepts achieving scores above 0.6. In contrast, 8M models show compressed distributions with medians below 0.1, indicating substantially weaker concept capture across the board.

\begin{figure}[h]
    \centering
    \includegraphics[width=0.9\linewidth]{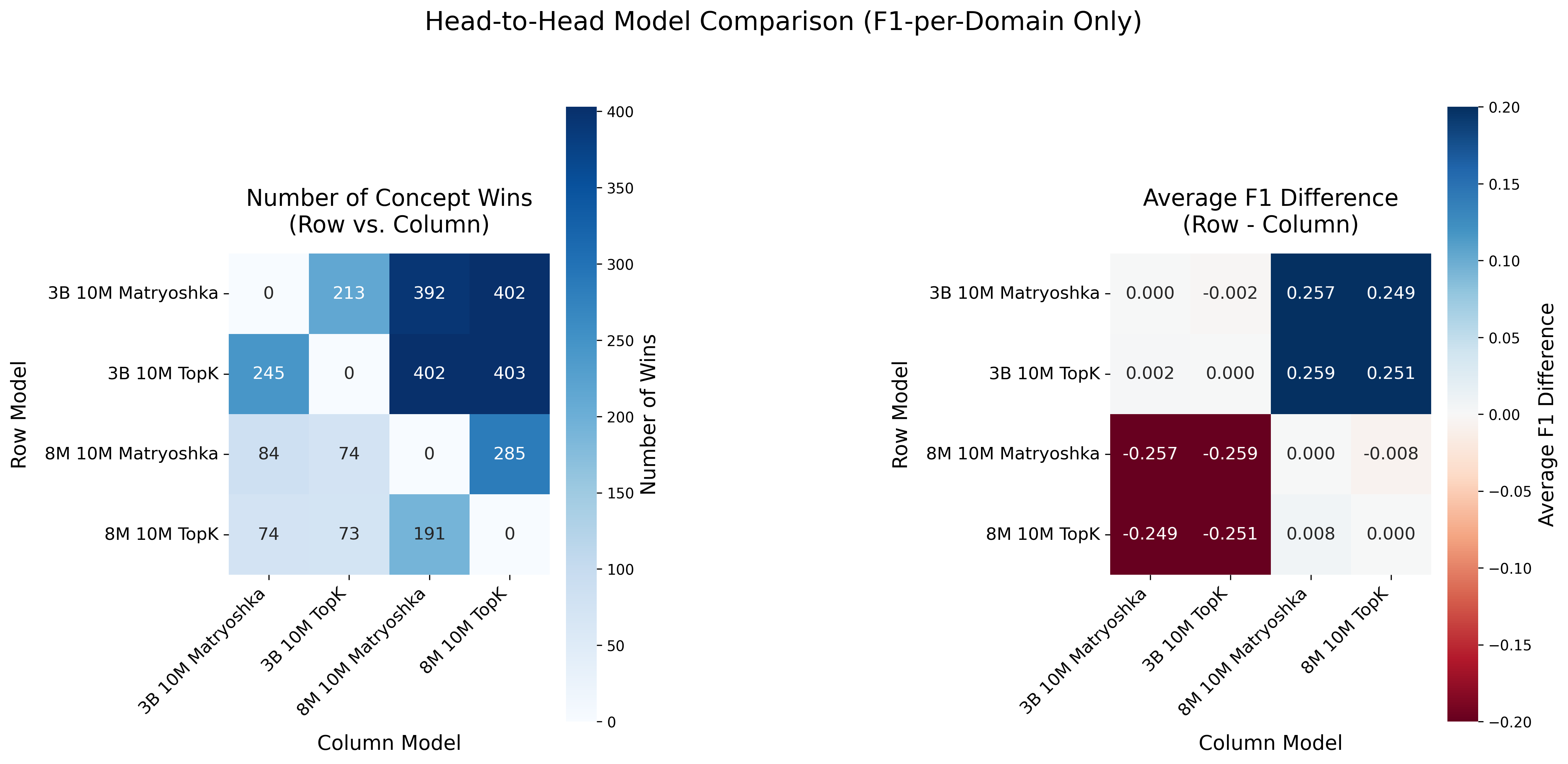}
    \caption{Left: Number of concepts where the row model achieves a higher maximum F1 score for a given concept than the column model. Right: Average score difference of the highest F1 scores for a given concept between models (row minus column) across all concepts. Each cell compares the best-performing feature for each concept between model pairs. Darker blue indicates stronger performance advantage for the row model.}
    \label{fig:f1_heatmap}
\end{figure}

The comparison heatmaps demonstrate relative performance between model variants by comparing their best-performing features for each concept. The left heatmap shows the number of concepts where each row model achieves a higher F1 score than the column model, while the right heatmap shows the average F1 score difference (Figure \ref{fig:f1_heatmap}). The 3B models outperform 8M variants on approximately 400 concepts, with an average F1 improvement of 0.25. Within each scale, architectural variants show minimal differences (average F1 difference $<$ 0.01), suggesting model scale rather than architecture drives concept capture ability.

\newpage

\section{Structure Prediction} 
\label{app:casp14_dataset}

\subsection{CASP14 Dataset}
For computational efficiency, we filtered the CASP14 dataset to exclude sequences longer than 700 residues to enable single-GPU evaluation. We further filtered out targets where ESMFold predictions showed RMSD $> 10$ Å compared to the experimental structure, resulting in 17 of the original 34 targets

\textbf{Data.} We downloaded the CASP14 targets dataset from the CASP14 website \href{https://www.predictioncenter.org/download_area/CASP14/targets/casp14.targets.T.public_11.29.2020.tar.gz}{here.}

After filtering, we benchmark on these targets: [T1024, T1025, T1026, T1029, T1030, T1032, T1046s1, T1046s2, T1050, T1054, T1056, T1067, T1073, T1074, T1079, T1082, T1090].

\subsection{Sparsity Sweep}
\begin{figure}[h]
    \centering
    \includegraphics[width=0.5\linewidth]{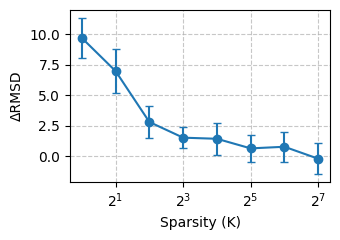}
    \caption{CASP14 $\Delta$RMSD Results Across Sparsity Levels $K$ for Matryoshka SAE Architecture}
    \label{fig:rmsd_k_sweep}
\end{figure}

We report the difference in RMSD on the above CASP14 dataset between ESMFold and SAE reconstructions across sparsity levels for Matryoshka SAEs. In Figure \ref{fig:rmsd_k_sweep}, we see that with only $2^3$ to $2^5$ active latents, SAEs can reasonably reconstruct structure prediction performance. These results raise questions on how much information is required per token for structure prediction.

\section{Contact Map Prediction} \label{app:cm}

\textbf{Motivation.} \citet{Ovc2024} introduced the Categorical Jacobian calculation to extract coevolutionary signals from protein language models in an unsupervised manner via masked language modeling. Given that SAEs also learn through unsupervised training, we adapt this approach for contact map prediction to evaluate whether SAEs preserve coevolutionary patterns and structural information learned by ESM2.

\begin{figure}[h]
    \centering
    \includegraphics[width=0.5\linewidth]{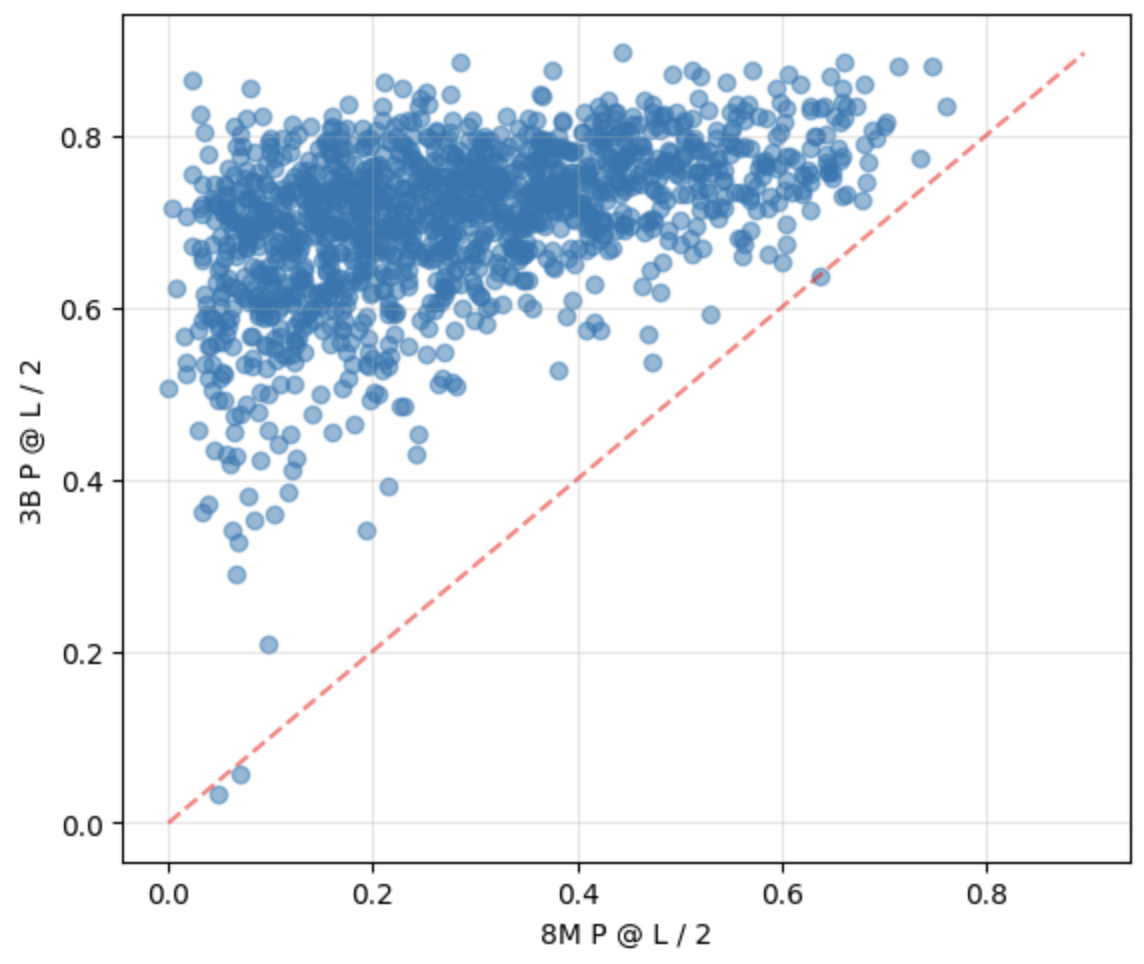}
    \caption{Contact Map Accuracy for Cat. Jacobian on ESM2 8M vs 3B, N=1412/1431}
    \label{fig:cm_whole}
\end{figure}

\textbf{Data.} We randomly sampled 50 proteins from the evaluation dataset in \cite{Ovc2024} due to computational constraints. As shown in Fig. \ref{fig:cm_whole}, our results replicate the authors' findings, demonstrating superior contact map accuracy across nearly the entire sample. Based on these consistent results, we believe our conclusions would generalize to the complete dataset.

\textbf{Hyperparameters.} In Fig \ref{fig:cm_plot}, our SAEs were trained on layer 36 with TopK architecture with expansion factors 16x for ESM2-8M (dict size 10240) and 8x for ESM2-3B (dict size 20480). We did not rerun for the Matryoshka architecture given TopK's similar language modeling reconstruction performance and how computationally expensive it is.

\newpage

\section{ESMFold Steering Case Study} \label{app:case_study}

\subsection{Feature Steering Background} \label{app:steering_background}

Following the method described in \cite{templeton2024scaling}, each target feature is represented by its corresponding decoder vector ($d_{(i)} = W_{dec}[i, :]$), which is normalized to unit norm during training, and an intervention is performed by adding a scaled version of this vector to the hidden state ($h_l \leftarrow h_l + \alpha \cdot d_{(i)}$) to amplify or suppress that feature's contribution. In our approach, we apply a maximum activation scaling factor during training and subsequently scale the encoder and decoder biases at inference. This means our effective steering coefficient, reported in all figures and results, is given by $\alpha' = \text{norm\_factor} \cdot \alpha$, ensuring consistent steering effects across different model configurations.

\subsection{Details of Feature Steering Case Study} \label{app:case_study_details}

\begin{figure}[h]
    \centering
    \includegraphics[width=0.5\linewidth]{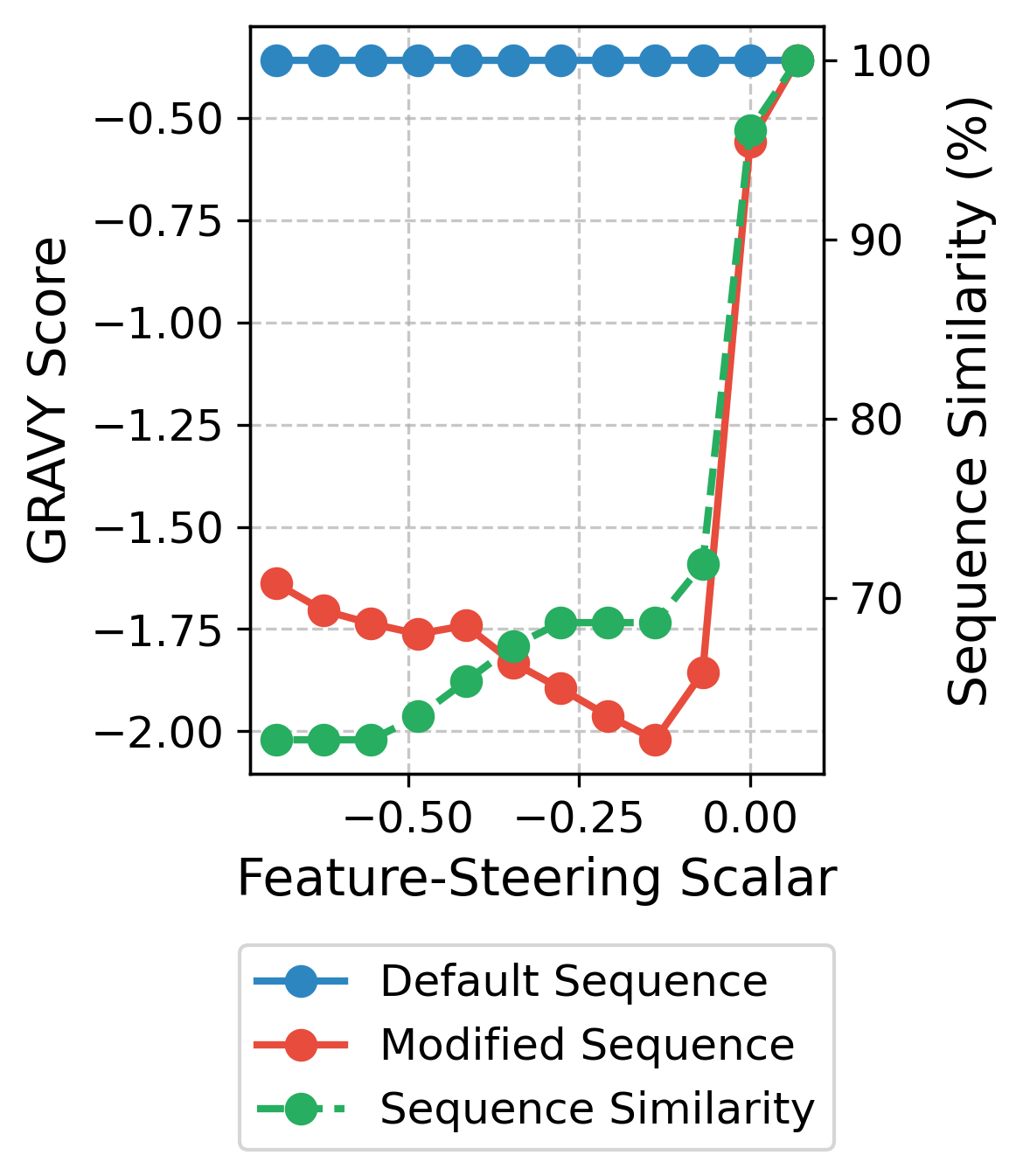}
    \caption{Plot of sequence GRAVY scores overlaid with sequence similarity with increased feature steering}
    \label{fig:gravy_seq_similarity}
\end{figure}

Our intervention framework employed two complementary approaches: (1) examining sequence-level changes through standard ESMFold predictions, and (2) directly intervening on hidden representations while maintaining the true input sequence in an ablated version that isolated intervention effects. This design allowed us to dissect how structural information flows through the model. The ablated model showed minimal performance differences (Figure \ref{fig:casp14_table}).

The feature (f/2) was identified in the first group of a Matryoshka SAE trained on hidden layer 36 (dictionary size 10,240, group fractions [0.002, 0.0156, 0.125, 0.8574], sparsity k=20), showing strong correlation with residue hydrophobicity. We selected the Matryoshka SAE architecture as its hierarchical encoding tends to capture broader concepts in earlier groups, which we hypothesized would enable more robust steering across diverse proteins. We applied the steering methodology from \cite{templeton2024scaling}, adding the decoder vector with coefficient $\alpha$ = -0.275. SASA was computed using FreeSASA \cite{Mitternacht_2016} with default parameters. The observed RMSD of 2.76 Å aligns with typical deviations for ESMFold with layer 36 LM head ablation, while ablating other ESM2 layer representations resulted in RMSD of 0.191 compared to regular ESMFold (Figure \ref{fig:casp14_table}). 

As shown in Figure \ref{fig:gravy_seq_similarity}, steering progressively decreased sequence GRAVY scores, indicating a shift toward hydrophilic residues. When modified sequences were processed through default ESMFold, we observed even larger SASA increases, consistent with replacing hydrophobic core residues with hydrophilic ones. This effect manifested through two distinct mechanisms: direct biasing of the structure module toward more exposed conformations (even with correct sequence information), and shifting sequence predictions toward more hydrophilic residues when allowed to influence the sequence module. Notably, structural changes persisted even when preserving the original sequence, suggesting ESMFold's predictions are influenced by hidden representations corresponding to the modified sequence despite having access to correct sequence information.

\subsection{Validation of Case Study Feature Steering Directionality and Specificity} \label{app:case_study_controls}
\begin{table}[htbp]
  \centering
  \begin{tabular}{lcccc}
    \toprule
    Scalar & Modified SASA & SASA Change & Modified GRAVY & GRAVY Change \\
    \midrule
    0.069 & 8369.55 & 0.00\% & -0.3588 & 0.00\% \\
    0.208 & 8351.58 & -0.21\% & -0.0595 & +83.42\% \\
    0.346 & 8568.43 & +2.38\% & 0.6569 & +283.08\% \\
    0.554 & 8616.49 & +2.95\% & 1.3046 & +463.60\% \\
    0.693 & 9719.28 & +16.13\% & 1.5255 & +525.17\% \\
    \bottomrule
  \end{tabular}
  \caption{Positive steering control results showing changes in SASA and GRAVY scores relative to baseline values.}
  \label{tab:positive_steering}
\end{table}

\begin{table}[htbp]
  \centering
  \begin{tabular}{llccc}
    \toprule
    Feature & Metric & \multicolumn{3}{c}{Scalar Values} \\
    \cmidrule(lr){3-5}
    & & -0.069 & -0.346 & -0.693 \\
    \midrule
    Feature 5 (Group 1) & $\delta$SASA & 0.00\% & +0.38\% & -0.54\% \\
    & $\delta$GRAVY & 0.00\% & +40.40\% & +52.34\% \\
    \midrule
    Feature 39 (Group 2) & $\delta$SASA & 0.00\% & 0.00\% & 0.00\% \\
    Feature 1381 (Group 3) & $\delta$SASA & 0.00\% & 0.00\% & 0.00\% \\
    Feature 7921 (Group 4) & $\delta$SASA & 0.00\% & 0.00\% & 0.00\% \\
    & $\delta$GRAVY & 0.00\% & 0.00\% & 0.00\% \\
    \midrule
    Original & $\delta$SASA & +128.44\% & +137.12\% & +112.89\% \\
    Experiment & $\delta$GRAVY & -417.34\% & -446.21\% & -356.69\% \\
    \bottomrule
  \end{tabular}
  \caption{Comparison of random feature controls with original experiment, showing percentage changes relative to unsteered (scalar=0) values.}
  \label{tab:random_features}
\end{table}

We performed positive steering control experiments to validate the directional behavior of our feature, applying scalar values from 0.069 to 0.693. As predicted, steering the feature positively resulted in a systematic increase in hydrophobicity (GRAVY score) while maintaining structural stability at moderate steering strengths. The modified sequence GRAVY scores increased from -0.36 to 1.53, reflecting enhanced hydrophobic content, while SASA values remained stable ($\leq$ 3\% change) until the highest scalar values where minor structural perturbations emerged.

Additionally, to assess the specificity of identified feature, we performed negative steering experiments on randomly selected features from each Matryoshka group (features 5, 39, 1381, and 7921). Unlike our target feature, random features showed minimal response to steering, with three features showing no changes ($\leq$ 0.1\% variation) across all metrics and one feature (Feature 5) showing only minor variations incomparable to the dramatic shifts observed in our main experiment.

\section{Implementation Details}

\subsection{SAE Implementation Details} \label{app:sae_details}
For each token of a protein sequence, a transformer encoder-based PLM outputs an embedding $x$ in each layer $\ell$. To learn meaningful representations, sparsity constraints are imposed on the latent vector $z$. This sparsity can be achieved either through L1 regularization terms in the loss function or through specialized activation functions that directly restrict the number of non-zero elements. In practice, the number of active (non-zero) elements $K$ satisfies $K \ll d \ll n$, preventing degenerate solutions such as the identity map.

\textbf{Training and Normalization.} During training, we normalize our embeddings to have average unit L2 norm, enabling hyperparameter transfer between layers. This normalization is reversed during inference by scaling the biases according to the procedure detailed in \citet{marks2024dictionary_learning}. This approach ensures consistent performance across different layers while maintaining the model's representational capacity.

\subsection{Code} \label{app:impl}
For all SAE implementations, we use \cite{marks2024dictionary_learning}. For our L1 regularized SAEs, we follow \cite{conerly2024aprilupdate}. For TopK, we follow \cite{gao2024scalingevaluatingsparseautoencoders}. In Fig \ref{fig:delta_ce}, we standardize hyperparameters with an expansion factor of 8x (dict size 20480) and batch size 2048 and use best learning rate.

\subsection{Data Curation} \label{app:data_curation}
Following the approach outlined in the InterPLM paper, we first process the UniRef50 FASTA file by iterating over each protein sequence. We remove any sequence that exceeds 1022 tokens, as ESM-2 cannot process longer inputs. From the remaining sequences, we randomly select a predetermined number of proteins to create our working dataset. This dataset is then divided into shards of 1000 sequences each, facilitating efficient handling during training.

\subsection{Optimizing the Dataset and Training Procedure}
After generating embeddings with ESM-2, the activations are stored as tensors representing entire shards. We then use Litdata's optimize function to iterate through each tensor and split it into fixed-size batches. The optimized dataset achieves speedup by reducing streaming I/O overhead to AWS S3 during training. Additionally, the optimized dataset uses Litdata's StreamingDataset and StreamingDataloader where we shuffle the batches during training, while also leveraging PyTorch Lightning for multi-GPU training support.

\newpage
\bibliographystyle{unsrtnat}
\bibliography{references}

\end{document}